\documentclass{article}

\usepackage{PRIMEarxiv}

\usepackage[utf8]{inputenc} 
\usepackage[T1]{fontenc}    
\usepackage{hyperref}       
\usepackage{url}            
\usepackage{booktabs}       
\usepackage{amsfonts}       
\usepackage{nicefrac}       
\usepackage{microtype}      
\usepackage{lipsum}
\usepackage{fancyhdr}       
\usepackage{graphicx}       
\graphicspath{{media/}}     

\usepackage{chemformula} 
\usepackage{textalpha}
\usepackage{siunitx}
\DeclareSIUnit\angstrom{\text {Å}}
\DeclareSIUnit\bar{\text {bar}}

\newcommand{\ie}{\emph{i.e.}}

\title{Epitaxial \ch{Sr(Sn, Ge)_xTi_{1-x}O3} buffer layers for continuous strain engineering on \ch{SrTiO3} substrates
}

\author{
  Ruben Hamming-Green\thanks{Joint first-authors}\\
  Zernike Institute for Advanced Materials\\ Faculty of Science and Engineering\\ University of Groningen\\ Groningen, The Netherlands\\
  CogniGron Institute, University of Groningen\\ Groningen, The Netherlands \\
  \texttt{r.p.hamming-green@rug.nl} \\
   \And
  Ewout van der Veer\footnotemark[1] \textsuperscript{\space}\thanks{Current affiliation: Faculty of Physics, University of Duisburg-Essen, Duisburg, Germany;
  Research Center Future Energy Materials and Systems (RC FEMS), Research Alliance Ruhr, Bochum, Germany\\} \\
  Zernike Institute for Advanced Materials\\ Faculty of Science and Engineering\\ University of Groningen\\ Groningen, The Netherlands\\
  \texttt{ewout.vanderveer@uni-due.de} \\
  \AND
  Beatriz Noheda \\
  Zernike Institute for Advanced Materials\\ Faculty of Science and Engineering\\ University of Groningen\\ Groningen, The Netherlands\\
  CogniGron Institute, University of Groningen\\ Groningen The Netherlands\\
  \texttt{b.noheda@rug.nl}
}

\begin{document}
\maketitle

\begin{abstract}
  Epitaxial strain plays a key role in determining the structure and functionality of thin films, with the choice of substrate being traditionally used to control the magnitude  of the applied strain.
  However, even in the large family of perovskite materials, this allows for only a limited, discrete set of  strain states to be achieved.  Here we report on an approach to controlling epitaxial strain for the growth of perovskite materials by involving a single \ch{SrTiO3} substrate (the most available perovskite in single crystal form) and a buffer layer that consists of the solid solution \ch{Sr(Sn, Ge)_xTi_{1-x}O3}, of which the lattice parameter can be tuned in a continuous fashion, from \qty{3.880}{\angstrom} up to \qty{4.007}{\angstrom}, while maintaining coherent epitaxial growth on \ch{SrTiO3} with high quality interfaces.
  Using a \ch{BaTiO3} overlayer as a model system, we show that changes to the buffer layer composition, \ie\ increase of in-plane  lattice parameter, change the strain state of \ch{BaTiO3} from fully relaxed, through highly compressively strained, to an exotic state showing 'inverted' epitaxy in which the buffer layer is relaxed from the substrate but lattice matched to the overlayer.

\end{abstract}


\section{Introduction}

One of the most ubiquitous methods for exploring the properties of epitaxial thin films is modifying their strain state.\cite{Pertsev88911, Choi2004, haeni2004, dieguez2005, ederer2005}
In the most interesting functional materials, and particularly in complex oxides such as perovskites, minute changes in the interatomic distances lead to substantial, and often drastic, changes in their magnetic, ferroelectric or (super)conducting properties.
Most typically, epitaxial strain engineering of a crystalline thin film is accomplished by selecting a crystalline substrate with a slightly different lattice parameter on which the target material is deposited.
The lattice parameters of the overlayer material then expand or contract to form a coherent interface with the substrate, creating a tensile or compressive strain in this layer, respectively.
The film, thus, grows under strain up to a certain critical thickness, above which the accumulated elastic energy becomes too large and the film relaxes by forming twin walls (in ferroelastic materials), dislocations or other defects.

There is a set of perovskite substrates that are commonly used for strain engineering of other functional perovskites and provide lattice parameters ranging between \qty{3.787}{\angstrom} (\ch{LaAlO3}) and \qty{4.02}{\angstrom} (\ch{PrScO3}).\cite{Uecker2008}
However, this discrete set of lattice parameters does not allow exact tuning to any desired strain state.
The are various techniques that enable continuous strain engineering, but these require large amounts of optimization, as the surface chemistry, growth parameters and substrate preparation are different for each substrate.
It is also possible to apply external, uniaxial strain, for example through use of a piezoelectric substrate.\cite{Herklotz_2010, hicks_piezoelectric-based_2014}
This has the advantage of allowing reversibility of the strain state but it requires electrodes to access the substrate electrically and limits the substrates that can be used.

Another common technique is the growth of a partially relaxed buffer layer between the film and the substrate, which can provide continuous strain modification by varying its thickness.\cite{Janolin_2014}
The challenge in this case is that strain relaxation only happens in a narrow range of thicknesses and some materials relax quite abruptly at the critical thickness.\cite{deng_strain_2021}
Too large a lattice mismatch or sudden relaxation involves the creation of dislocations or point defects.
Avoiding these is particularly important for optimizing the transport properties of thin films\cite{Guo2020}, to avoid spurious magnetic phases and, in the case of ferroelectric thin films, to achieve a good endurance upon electric field cycling, which is largely dependent on a lack of defects in the structure.\cite{eom_fabrication_1993}
An additional challenge is that there are few materials that are highly conductive while also preserving high quality epitaxy, with \ch{SrRuO3} and \ch{La_{0.67}Sr_{0.33}MnO3} being some of the few that can function as a back electrode without reducing film quality.\cite{Koster2012, abuwasib_contact_2015, vailionis2011}
Therefore, a wider choice of substrates or buffer layers is required to be able to select the lattice mismatch more accurately.

Here we propose a simple system that allows for selection of in-plane lattice parameters by varying the composition, using \ch{SrTiO3} as a substrate, which is interesting for the epitaxial growth of many different materials.
By creating the solid solutions \ch{Sr(Sn, Ge)_xTi_{1-x}O3} (SSGTO), it is possible to create crystalline films with lattice parameters larger or smaller than those of \ch{SrTiO3}, and to continuously tune their lattice parameters. When used as buffer layers on \ch{SrTiO3} substrates, these allow for the strain in the overlying thin films to be precisely selected, and produce high quality films that require little additional growth optimization.

Strontium stannate (SSO) is a wide-bandgap perovskite material with a pseudocubic lattice parameter of \qty{4.03}{\angstrom}.\cite{Liu2011}
The wide bandgap and high electron mobility of alkaline earth perovskite stannates \ch{(Ba,Sr,Ca)SnO3} has attracted researchers to this material for its potential as a transparent electrode,\cite{Ismail-Beigi2015} perhaps to replace indium-tin-oxide,\cite{Prakash2019} and for use as a channel material in high-power transistors.\cite{Chaganti2020}
SSO has been demonstrated to grow epitaxially on STO substrates using atomic layer deposition (ALD)\cite{chen2019}, pulsed laser deposition (PLD) \cite{Alves2010, Liu2011} and molecular beam epitaxy (MBE) \cite{Raghavan2016}, though often these films exhibit high surface roughness.
The solid solution of \ch{SrSnO3} and \ch{SrTiO3} (SSTO) has also been previously explored: bulk, ceramic SSTO obeys Vegard's law\cite{Vegard1921}, \ie\ its lattice parameter varies linearly with the relative Sr/Ti content, between the bulk \ch{SrSnO3} (\qty{4.030}{\angstrom}) and \ch{SrTiO3} values (\qty{3.905}{\angstrom}).\cite{Singh2007, Stanulis2011}
Nevertheless, atomically flat SSO on STO has not been reported in the literature.

Unlike the materials previously discussed, \ch{SrGeO3} does not have a perovskite structure under ambient conditions.
Instead, it contains rings of \ch{GeO4}, rather than octahedra, and it is monoclinic (space group \emph{C2/c}). \cite{nishi_strontium_1997}
The high pressure phase, however, can be stabilized into a cubic perovskite structure with a lattice parameter of \qty{3.798}{\angstrom}, smaller than that of STO.\cite{nakatsuka_nakatsuka_sgo_2015, mizoguchi_germanate_2011}
The solid solution of \ch{SrGe_xTi_{1-x}O3} (SGTO) has not been previously reported.
Here we show that thin films of this compound can be stabilized in the perovskite phase when epitaxially strained to a \ch{SrTiO3} substrate.

The choice of strontium titanate as a substrate material is also important.
In addition to having a lattice parameter of \qty{3.905}{\angstrom}, which is near to a wide range of perovskite materials, it can also be grown epitaxially on Si-111\cite{Tambo1998, Saint-Girons2016}, making it a gateway material for bringing epitaxial oxides into industrial use.
Since many of the unique effects observed in oxides couple to strain, a buffer layer with variable lattice parameters that can be grown on such STO-buffered Si allows for a range of interesting and desirable properties of perovskite materials to be achieved on an industrially compatible backbone.

In this paper, we demonstrate the growth of Sn-substituted and Ge-substituted \ch{SrTiO3}, with Sn atomic content ranging from 0 to 1 and Ge atomic content ranging from 0 to 0.4, epitaxially on \ch{SrTiO3} substrates with atomically flat interfaces and surfaces.
The lattice parameters of the resulting buffer layers can be continuously tuned within an unprecedented large range from \qty{3.880}{\angstrom} to \qty{4.005}{\angstrom}.
To demonstrate the utility of this buffer layer for strain engineering, we employ a model system consisting of a ferroelectric barium titanate (\ch{BaTiO3}) overlayer grown on top of the buffer layer.
We subsequently characterize this overlayer using temperature-dependent x-ray diffraction (XRD) to measure its strain-dependent ferroelectric transition temperature, as well as visualizing the strain across the buffer layer and overlayer directly using atomic-resolution scanning transmission electron microscopy (STEM).

\section{Results}
\begin{figure}
  \centering
  \includegraphics[width=0.5\textwidth]{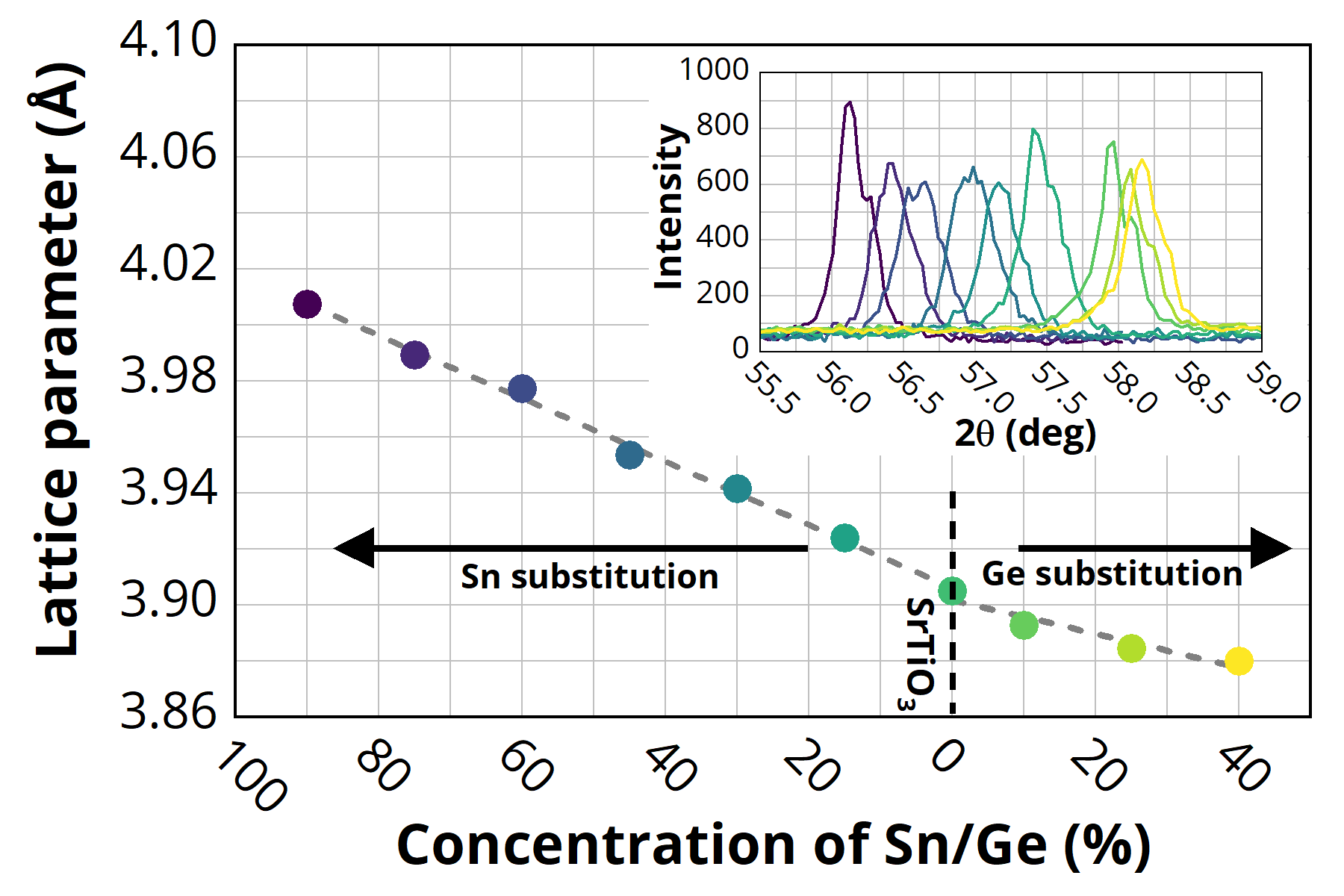}
  \caption{ \raggedright
    Lattice parameter of bulk \ch{SrSn_xTi_{1-x}O3} (SSTO) and \ch{SrGe_xTi_{1-x}O3} (SGTO) as a function of tin and germanium concentration.
    The inset shows the corresponding x-ray diffraction patterns around the (211) peak for all compositions.
  }
  \label{fig:bulk-xrd}
\end{figure}

Ceramic pellets of \ch{Sr(Sn, Ge)_xTi_{1-x}O3} (SSGTO) were produced by solid state synthesis (see section \ref{sec:experimental}), with $x =0.9, 0.75, 0.6, 0.45, 0.3, 0.15$ for the tin-containing targets and $x = 0.1, 0.25, 0.4$ for the germanium-containing targets. 
Powder XRD of the bulk SSGTO pellets, as seen in Figure \ref{fig:bulk-xrd}, shows that increasing the tin ($0 < x \le 1$) or germanium ($0 < x < 0.4$) concentration linearly modifies the lattice parameter of the SSTO between the bulk STO and SSO values.
Doping STO with tin results in a linear expansion of the lattice parameter due to the larger ionic size of \ch{Sn^{4+}} (\qty{69}{\pico\meter}\cite{shannon_revised_1976}) compared to \ch{Ti^{4+}} (\qty{60.5}{\pico\meter}\cite{shannon_revised_1976}), as earlier reported.\cite{Singh2007}
Conversely, doping with germanium (\ch{Ge^{4+}}, \qty{53}{\pico\meter}\cite{shannon_revised_1976}) causes a contraction of the lattice parameter.
The maximum expansion demonstrated here is \qty{2.6}{\percent} with respect to pure STO, for pure \ch{SrSnO3} (\ie\ SSTO with $x = 1$), while the maximum contraction is \qty{0.64}{\percent}, for SGTO with $x = 0.4$.

These pellets were then used as PLD targets for the growth of thin films. 
Firstly, thin films of SSTO were grown on STO(001) substrates, as described in section \ref{sec:experimental}.
The same growth parameters were used for all compositions, and the same number of laser pulses was used for each buffer layer.
On top of each layer, BTO films were deposited, again with the same growth parameters (see section \ref{sec:experimental}) and number of pulses in each case.

BTO was chosen as a good model system to test the effect of strain, as it has bulk lattice parameters of $a=b=\qty{3.994}{\angstrom}$, $c=\qty{4.03}{\angstrom}$, which are near the range of parameters being studied.
Additionally, it has been widely reported that epitaxial strain modifies both the lattice parameters and the Curie temperature of BTO,
which in bulk occurs at $T_\text{C} = \qty{130}{\degreeCelsius}$, together with a cubic-to-tetragonal phase change, but in epitaxial BTO thin films under high compressive strain, it has been shown to dramatically increase to well above \qty{500}{\degreeCelsius}.\cite{Choi2004}
Using other chemical pressure methods, such as integration of \ch{BaTiO3} with pure BaO phases, this transition temperature has even been pushed above \qty{1000}{\degreeCelsius}.\cite{Wang2021} The orientation of BTO, with the c-axis either in-plane or out-of-plane is also controllable through strain modification by selecting substrates closer to the a- or c- crystal axis, or can even be coaxed into different orientations by growing at high temperatures on substrates with thermal expansion mismatch.\cite{srikant1995}

\begin{figure}
  \centering
  \includegraphics[width=0.9\textwidth]{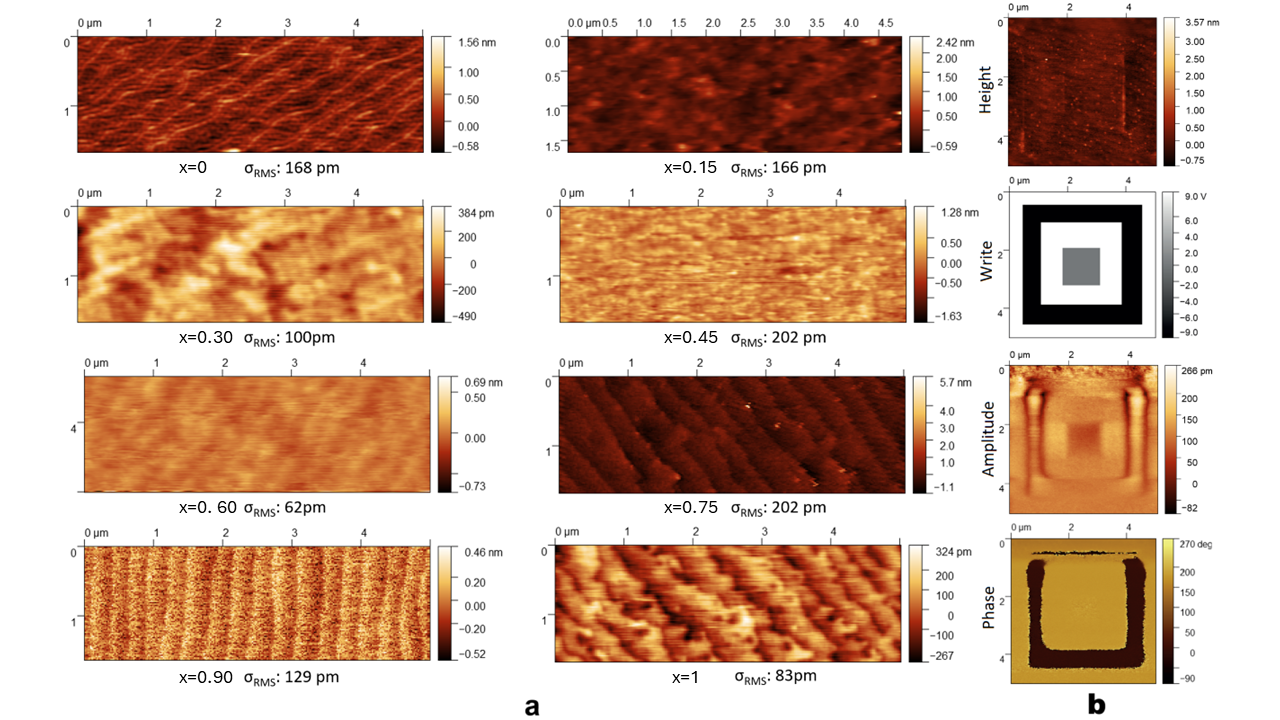}
  \caption{ \raggedright
    a) AFM images of \ch{BaTiO3} films grown on buffer layers of \ch{SrSn_xTi_{1-x}O3} with various concentrations.
    The composition of the buffer layers and the roughness of the heterostructure surface are denoted under each image.
    b) Piezoresponse force microscopy (PFM) on a BTO/SSTO film, with $x = 0.75$. From top to bottom: Topography (height) image of the selected area; Written box-in-box pattern with alternating positively and negatively polarized squares and maximum amplitude of 9 V; PFM amplitude signal of the same area; and at the bottom, PFM image of the same area.  The images are  consistent with a upward intrinsic polarization and a switching voltage between 6 and \qty{7}{\volt}.
  }
  \label{fig:afm}
\end{figure}

The resulting thin films are shown to be of very high quality, despite their relatively large thickness of around 110-120 nm, and that they include two sublayers, as observed in Figure \ref{fig:afm}a.
The films have atomically flat surfaces, with most of them still showing the atomic terraces from the underlying substrate.
RHEED oscillations (see section \ref{sec:experimental}) are seen throughout the deposition process for both the SSTO/SGTO with Sn content below \qty{75}{\percent} and BTO layers in all samples, indicating highly controlled layer-by-layer growth.
Step-flow growth was achievable for films with Sn concentrations of \qty{75}{\percent} and above, indicating that high mismatch between the buffer layer and substrate (allowing the buffer layer to fully relax), combined with minimal mismatch between the lattice parameters of SSTO and BTO, may promote higher growth quality.

Piezoresponse force microscopy (PFM) of out-of-plane oriented BTO (Figure \ref{fig:afm}b) shows that the as-grown film has an out-of-plane polarization pointing towards the film surface, which can be switched by the application of a \qty{9}{\volt} tip bias.

\begin{figure}
  \centering
  \includegraphics[width = 0.8\textwidth]{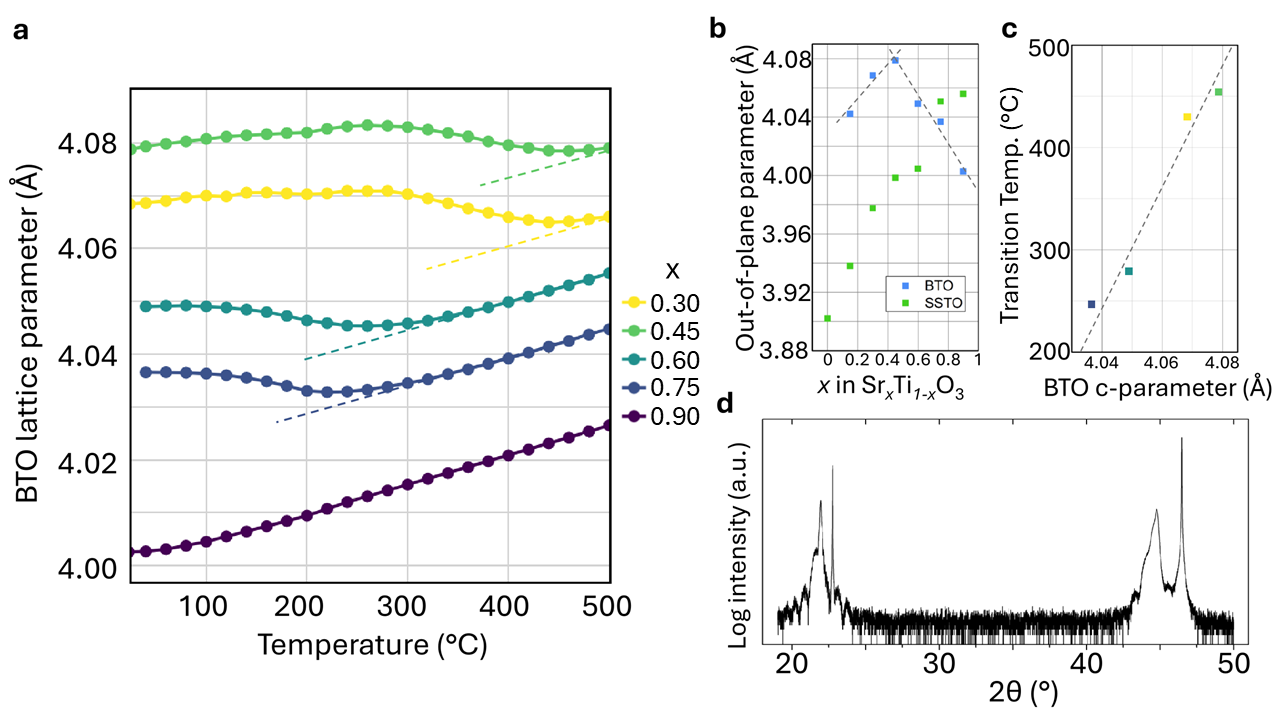}

  \caption{\raggedright
    a) BTO out-of-plane lattice parameter as a function of temperature for different SSTO buffer layer compositions. The dashed lines show the thermal expansion in the paraelectric phase, following the data from the x= 0.90 composition.
    The phase transition temperature $T_\text{C}$ increases with increasing compressive strain. 
    b) Out-of-plane lattice parameter at room temperature for different buffer layer compositions. 
    An opposite trend is observed in the BTO lattice parameters for $x < 0.5$ and $x > 0.5$. 
    c) Transition temperature $T_\text{C}$ \textit{versus} out-of-plane lattice parameter of BTO.
    d) A $2\theta - \omega$ scan of a BTO film on \ch{SrSn_{0.75}Ti_{0.25}O3} on STO.
    }
  \label{fig:tc}
\end{figure}

To test whether the unit cell of the SSTO had an effect on the $T_\text{C}$ of the BTO, the films were mounted on a heating stage and $2\theta - \omega$ XRD scans were conducted up to \qtyrange{500}{600}{\degreeCelsius}.
Samples were heated at \qty{10}{\degreeCelsius\per\minute}, and temperatures at which the scans were taken, were allowed to stabilize for \qty{2}{\minute} before each scan.
The peak positions of STO, SSTO and BTO peaks at each temperature were determined by fitting pseudo-Voigt curves to the XRD patterns.
$T_\text{C}$ for each case was determined using the intersection of the linear fit of the lattice parameter \emph{vs.} temperature curve above the phase transition (with positive slope)  with a linear fit of a few points below the transition (negative slope) (see Figure \ref{fig:tc}).

Figures \ref{fig:tc}a,b show that for $x \ge 0.45$, the out-of plane lattice parameter of the BTO overlayer decreases with increasing x, while the opposite is true for $x < 0.45$. 
This suggests that the BTO film is fully strained to the SSTO buffer layer above $x = 0.45$, while at lower x, the BTO layer starts to relax, \ie the buffer layer is no longer able to impose its lattice parameter over the full thickness of the BTO overlayer.  
Nevertheless, the out-of-plane lattice parameter increases monotonically with the Sn content of the buffer layer, similar to the behavior in the bulk pellets (Figure \ref{fig:bulk-xrd}).
As expected, the Curie temperature $T_\text{C}$ is linearly correlated to the out-of-plane lattice parameter of the BTO layer (Figure \ref{fig:tc}c).
Scanning electron microscopy experiments were performed in order to further clarify the relaxation behavior in the SSTO buffer layer and the BTO overlayer, and to determine the way in which the epitaxial strain is accommodated in these films (see next section). 
Temperature-dependent XRD tests were not performed on the samples grown on SGTO due to the even larger lattice mismatch.

\begin{figure}
  \centering
  \includegraphics[width=\textwidth]{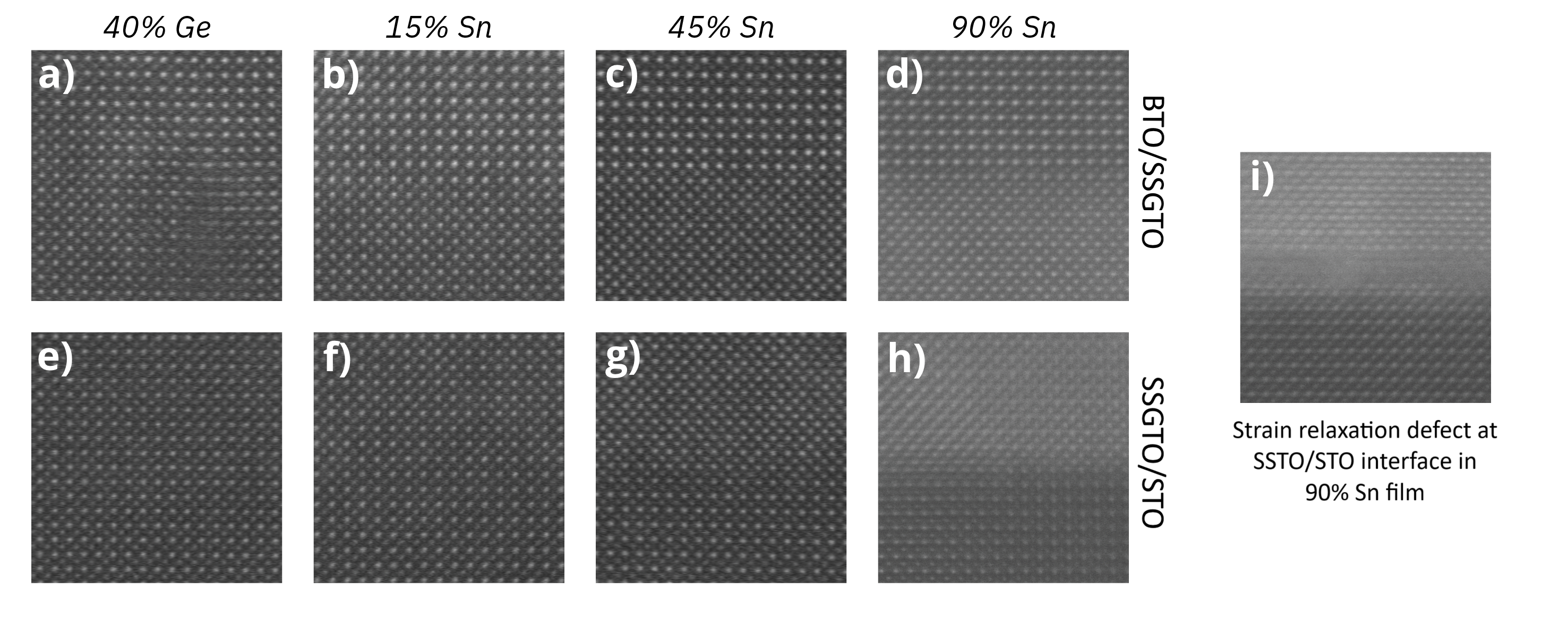}
  \caption{ \raggedright
    STEM images of the a-d) BTO/SSGTO and e-h) SSGTO/STO interfaces for four compositions of the SSGTO buffer layer.
    The films with intermediate compositions $x = 0.15$ and $x = 0.45$ show coherent epitaxial growth at both interfaces.
    The SGTO film shows coherent growth at the SGTO/STO interface but defective and highly strained growth at the BTO/SGTO interface.
    Conversely, the SSTO film with $x = 0.9$ has a coherent BTO/SSTO interface and an SSTO/STO interface with regions of coherent growth and regions where the strain is relaxed through interfacial defects (i).
    In both cases, the appearance of these defects can be understood by considering the lattice mismatch between respective layers.
  }
  \label{fig:filtered-stem}
\end{figure}

\begin{figure}
  \centering
  \includegraphics[width=0.8\textwidth]{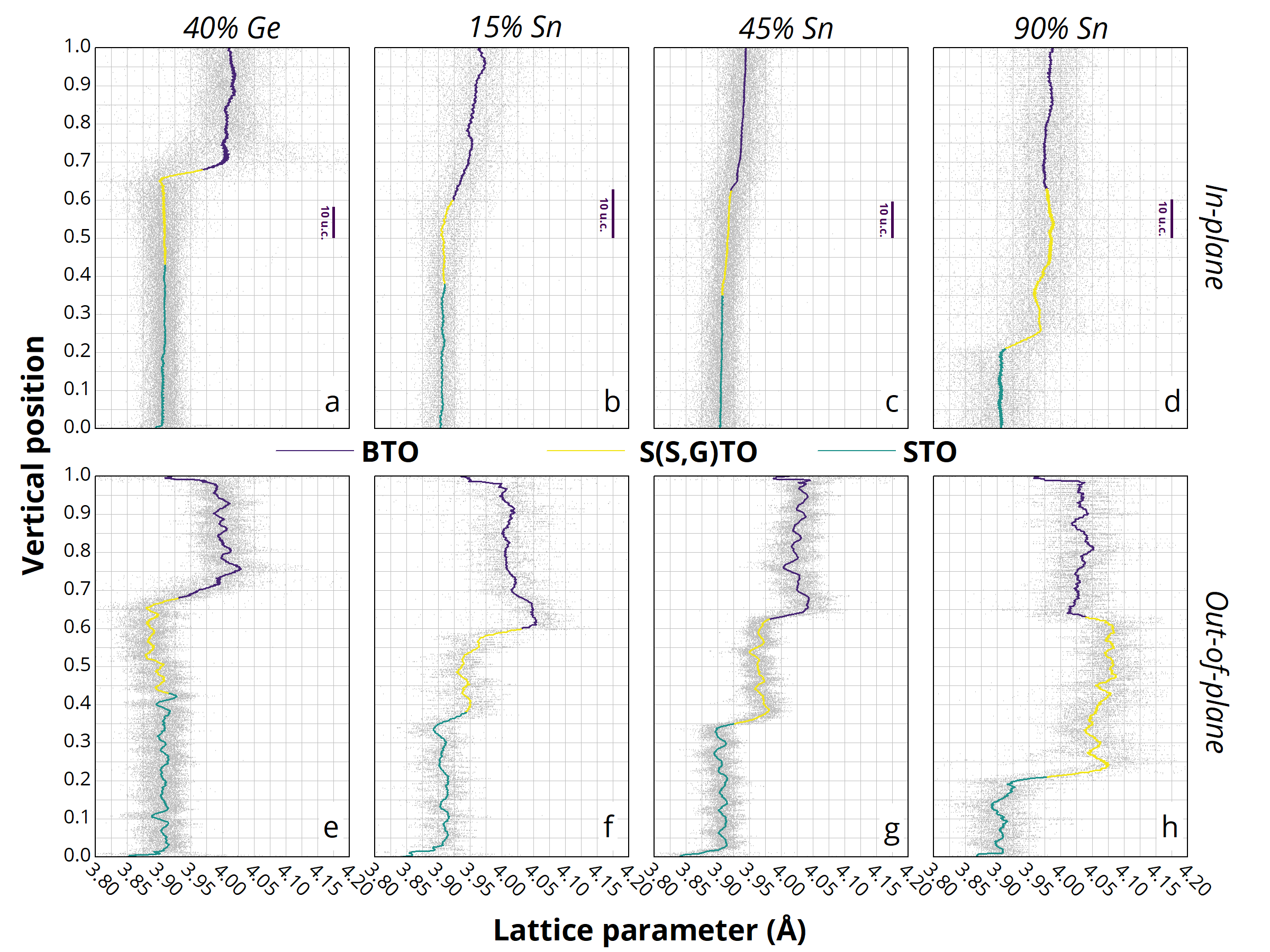}
  \caption{ \raggedright
    Profiles of the out-of-plane (a-d) and in-plane (e-h) lattice parameters as a function of relative depth in the film as calculated using STEMfit.
    The vertical axis shows the relative vertical position in the image as shown in Figure \ref{fig:filtered-stem}, \ie\ 1 corresponds to the top of the image, 0 to the bottom.
    The light gray markers show the local lattice parameter value of each atomic column detected in the image.
    The colored line represents a moving average of this data along the vertical axis.
    The different colors of this line correspond to the different layers of the film, as indicated in the legend.
    The length of each of the scale bars in (a-d) represents 10 unit cells in each corresponding image.
  }
  \label{fig:lp_profile}
\end{figure}

Figure \ref{fig:filtered-stem} shows filtered STEM images of the SSTO samples with $x = 0.1$, $x = 0.55$ and $x = 0.85$ and of SGTO with $x = 0.4$.
The out-of-plane and in-plane lattice parameter profiles obtained from analysing these images (see Experimental methods section) are shown in Figure \ref{fig:lp_profile}.
The sample containing Ge with $x = 0.4$ (a,e), which has the smallest bulk SSGTO lattice parameter, has its SSGTO layer fully strained to the underlying STO substrate.
The BTO layer grown on top is fully relaxed due to the large lattice parameter mismatch between SSGTO and BTO.
Relaxation in this sample occurs mostly through defects at the BTO/SSGTO interface (see Figure \ref{fig:filtered-stem}a).
Extended areas of coherent growth are nevertheless present.

Increasing the SSGTO lattice parameter (Figure \ref{fig:lp_profile}b,f), reduces the lattice mismatch, now allowing BTO to be strained to SSGTO.
The in-plane strain is gradually relaxed, mostly in the BTO layer, reaching a value of approximately \qty{3.975}{\angstrom} at a thickness of approximately \qty{12}{\nano\meter}.
Upon further increasing the Sn content of the buffer layer to $x = 0.45$ (Figure \ref{fig:lp_profile}c,g), the BTO becomes more coherent over its entire thickness, showing only a slight relaxation of the in-plane lattice parameters from \qty{3.930}{\angstrom} to \qty{3.943}{\angstrom}.
This observation is consistent with the fact that this buffer layer composition has the largest out-of-plane lattice parameter and $T_\text{C}$ (Figure \ref{fig:tc}).
Indeed, at this crossover composition ($x = 0.45$), the lattice parameter of the buffer layer is nearly exactly intermediate between that of the STO substrate and the in-plane lattice parameter of the BTO overlayer.
As such, this composition of SSTO is ideal for matching the lattice parameters of STO and BTO, providing optimal transfer of strain from the substrate to the overlayer. 

Finally, in the sample with the buffer layer with the largest lattice parameter ($x = 0.9$,  Figure \ref{fig:lp_profile}d,h), the lattice mismatch between SSGTO and the STO substrate is so large, that the former becomes fully relaxed from the substrate.
The strain is relaxed through interfacial defects at the STO/SSGTO interface, as seen in Figure \ref{fig:filtered-stem}d. However, the BTO layer does grow fully strained to the SSTO buffer.
The out-of-plane lattice parameter of the SSTO still remains larger than the in-plane lattice parameter (\ie\ the structure is not cubic), suggesting that the buffer layer is now epitaxially strained to the BTO overlayer.
As can be seen in the STEM images in Figure \ref{fig:filtered-stem}, despite being strained to the BTO layer, some degree of coherency is, nevertheless, maintained between the STO substrate and the SSTO buffer layer. 
These results suggest that the SSTO buffer layer only relaxes from the STO substrate after the BTO overlayer has been deposited (either during BTO deposition or during cooling), at which point it becomes strained to the overlayer in a kind of non-standard, 'inverted' epitaxy. 

From the analysis of the STEM images, we can also see that the out-of-plane lattice parameter displays complementary behavior to the in-plane lattice parameter, though with more random variation due to sample drift during the recording of the STEM images.
In the fully relaxed BTO layer on SGTO (Figure \ref{fig:lp_profile}e) the out of plane lattice parameter shows a continuous and rapid increase at the interface with the SGTO, quickly reaching the bulk lattice parameter of \qty{4.00}{\angstrom}.

For the SSTO samples with $x = 0.15$, the BTO out-of-plane lattice parameter increases abruptly at the interface with the SSTO to reach a value of \qty{4.05}{\angstrom} for the first few unit cells.
This is different from the gradual change observed in the in-plane lattice parameters and it could indicate an accumulation of oxygen vacancies near the interface to assist the epitaxial growth.

The sample with $x = 0.45$ does not show this interfacial out-of-plane expansion of the BTO overlayer.
Instead, the out-of-plane lattice parameter of the BTO layer in this case is nearly constant, again showing that this overlayer is optimally strained in-plane by the combination of STO substrate and SSTO buffer layer. 

In summary, below the crossover point $x = 0.45$, the SSTO buffer layer is fully strained to the STO substrate, but its lattice parameter is too small for the BTO overlayer to remain strained to it, leading to a decrease of the out-of-plane lattice parameter of BTO and, concomitantly, of its Curie temperature.
Above the crossover composition of the SSTO, the BTO layer is fully strained to the buffer layer, but the lattice parameter of the latter is larger, so the amount of epitaxial strain it imposes on the overlayer is smaller.
Exactly at the crossover, the strain transfer by the buffer layer is optimal, so the effect on the out-of-plane lattice parameter and $T_\text{C}$ is at a maximum.

\section{Conclusion}
Reliable, easy control over biaxial, epitaxial strain in thin films is highly desirable and necessary to fine tune many of the most technologically intriguing effects in complex oxide thin films.
Many of the standard methods for controlling strain are non-ideal.
They are either non-continuous (such as selecting substrates with different lattice parameters, which also often require different treatment steps), require a large amount of optimization (in the case of strain-releasing buffer layers), or rely on changing the layer under investigation itself (modifying the thickness, stoichiometry, or growth parameters).
During epitaxial growth, several growth parameters are strongly interdependent such that changing the strain state also implies changing other material properties. 
In order to isolate the effect of strain, it is therefore desirable to have a strain tuning method that relies on the change of a single growth parameter. 
We show that by changing composition of the buffer layer alone, we are able to achieve similar quality films with the same growth conditions.
SSTO on STO is a promising candidate for this purpose, as its lattice parameter is linearly proportional to the Ti concentration, it can be grown as thin films with atomically flat surfaces and interfaces, and it can be integrated into the existing STO-buffered Si platform, giving it clear industrial potential.
The range of possible lattice parameter values (between \qty{3.90}{\angstrom} and \qty{4.03}{\angstrom}), which is widely interesting for the growth of perovskites, but also other materials of recent interest, such as \ch{HfO2}-based ferroelectrics, previously was accessible only through the use of scandate substrates.
Though scandates are ideal substrates for highly controlled epitaxial growth due to their low dislocation densities\cite{Uecker2008}, these materials are expensive and not readily available because of their reliance on rare earth elements.

While the region with lattice parameters larger than STO has been previously explored, in this work we propose that the region with lattice parameters smaller than those of STO is also accessible, through germanium substitution in STO.
BTO films grown with SGTO buffer layers also grow with good orientation and quality on STO substrates using identical growth parameters as those on SSTO buffer layers.
However, these BTO films are relaxed from the SGTO buffer layer due to the large lattice mismatch.
As such, BTO is no longer a good model system for this range, as the lattice parameter is too small for coherent growth, but we foresee that it will provide an interesting template for the growth of several other perovskites, such as \ch{PbTiO3}, \ch{SrTiO3}, \ch{KTaO3}, \ch{RENiO3} or \ch{(La,Sr)MnO3}, under compressive strain.

\section{Experimental Section}
\label{sec:experimental}
PLD targets were synthesized by solid state synthesis from strontium carbonate (\ch{SrCO3}, \qty{99.99}{\percent}, Alfa Aesar), titanium dioxide (\ch{TiO2}, \qty{99.6}{\percent}, Alfa Aesar), tin oxide (\ch{SnO2}, \textgreater \qty{99}{\percent}, Lamers \& Indemans) and germanium oxide (\ch{GeO2}, \qty{99.9999}{\percent}, Alfa Aesar).
The reagents were dried in vacuum overnight, then mixed in stoichiometric amounts to form \ch{Sr(Sn, Ge)_xTi_{1-x}O3}, hereafter named SSGTO, where $x = 0.9, 0.75, 0.6, 0.45, 0.3, 0.15$ for the tin-containing targets and $x = 0.1, 0.25, 0.4$ for the germanium-containing targets.
The mixtures were ball-milled at \qty{200}{rpm} for \qty{1}{\hour} in an agate mortar, then calcined at \qty{1400}{\degreeCelsius} for \qty{12}{\hour}.
The resulting powders were pressed into \qty{2}{\centi\meter} diameter pellets under \qty{0.77}{\giga\pascal} using poly(vinyl alcohol) as a binder.
The pellets were heated to \qty{300}{\degreeCelsius} for \qty{2}{\hour}, then sintered at \qty{1400}{\degreeCelsius} for \qty{12}{\hour}.
The pellets were then sanded to obtain a flat surface.
The \ch{BaTiO3} target was a polished single crystal (MaTecK).
These BTO/SSGTO stacks were grown on \ch{SrTiO3} (001) substrates (CrysTec GmbH), which were pre-treated to produce atomically flat \ch{TiO2} terminated terraces using a \qty{30}{\second} buffered hydrofluoric acid treatment followed by annealing at \qty{955}{\degreeCelsius} under \ch{O2} flow in a tube furnace.\cite{koster_sto_treatment_1998}

The SSGTO films were grown by pulsed laser deposition at \qty{650}{\degreeCelsius} under \qty{0.3}{\milli\bar} \ch{O2} pressure, with a fluence of \qty{1.9}{\joule\per\centi\meter\squared} and a laser frequency of \qty{0.3}{\hertz}.
A narrow growth temperature window was found as the most likely cause of the high film roughness observed in literature.
Square-like features are commonly reported if the growth temperature was too low (below \qty{600}{\degreeCelsius}), while increasing roughness is associated with higher temperatures above \qty{670}{\degreeCelsius}.
The BTO layer was then grown using the same fluence, but at \qty{630}{\degreeCelsius} and \qty{0.2}{\milli\bar} \ch{O2} pressure.
A frequency of \qty{1}{\hertz} was used for the first 200 pulses in order to form a seed layer,
The growth frequency was then increased to \qty{2}{\hertz} for the remainder of the film.
The samples were then annealed for \qty{30}{\minute} under \qty{300}{\milli\bar} of \ch{O2} pressure and cooled at \qty{5}{\degreeCelsius\per\minute} to room temperature.

Lamellae for electron microscopy were prepared of the SSTO films with $x = 0.10$, $x = 0.45$ and $x = 0.85$ as well as the SGTO films with $x = 0.40$ by focused ion beam milling using a gallium ion beam on an FEI Helios G4 CX FIB/SEM system.
They were imaged by high angle annular dark field scanning transmission electron microscopy (HAADF-STEM) on a Thermo Fisher Scientific Themis Z probe-corrected S/TEM at \qty{300}{\kilo\volt}.

Powder X-ray diffraction patterns of the SSGTO targets were collected to find their bulk lattice parameters.
The measurements were performed on a Bruker D8 Advance diffractometer in Bragg-Brentano geometry using Cu K\textalpha\ radiation.
All diffraction patterns show predominantly the cubic perovskite structure in space group Pm$\overline{3}$m.
Voigt functions were fitted to the (211) peaks of the diffraction patterns to accurately determine their positions, from which the lattice parameter were subsequently calculated.
The thin films were analyzed in a parallel-beam geometry using a Panalytical X'Pert Pro MRD multiplurpose X-Ray diffractometer.
XRD measurements were conducted in air as a function of temperature, while heating, by gluing the samples with silver paste to a ceramic plate and mounting them in an Anton Paar DHS1100.

STEM image analysis was performed using an in-house developed software package\cite{van_der_veer_stemfit_2023} written in the Julia language.\cite{bezanson_julia_2017}
In the first step, noise was removed from the images using a filtering procedure based on truncated singular value decomposition and gaussian convolution.
The images are then binarized using a Niblack local thresholding algorithm.\cite{niblack_introduction_1986}
The centroids of contiguous above-threshold regions are taken to be the positions of the atomic columns.
Next, an appropriate unit cell for the lattice is determined by finding the average vectors between pairs of neighboring atomic columns.
These vectors are used to define an affine transformation by mapping the atomic column positions in pixel units to positions in terms of the unit cell, facilitating further analysis and visualization.
Lattice parameter maps were constructed by calculating the average in-plane and out-of-plane distances of each atom to its neighbors in the transformed space.
For every image, the pixel size was determined using the unit cell size in the region of the image corresponding to the STO substrate and the known bulk lattice parameter of STO of \qty{0.3905}{\nano\meter}.

\section*{Acknowledgments}
The authors would like to acknowledge the financial support of the CogniGron research center and the Ubbo Emmius Funds (Univ. of Groningen).

\bibliographystyle{unsrt}  
\bibliography{references}

\end{document}